\documentclass[twocolumn,preprintnumbers,amsmath,amssymb]{revtex4}
\usepackage{graphicx}
\usepackage{dcolumn}
\usepackage{bm}

\begin{document}


\title{Fidelity approach to the disordered quantum XY model}

\author{Silvano Garnerone}
\email{garneron@usc.edu}
\affiliation{Department of Physics and Astronomy, University of Southern California, Los Angeles, CA 90089}
 
\author{N. Tobias Jacobson}
\affiliation{Department of Physics and Astronomy, University of Southern California, Los Angeles, CA 90089}

\author{Stephan Haas}
\affiliation{Department of Physics and Astronomy, University of Southern California, Los Angeles, CA 90089}

\author{Paolo Zanardi}
\altaffiliation[Also at ]{Institute for Scientific Interchange, Viale Settimio Severo 65, I-10133 Torino, Italy}
\affiliation{Department of Physics and Astronomy, University of Southern California, Los Angeles, CA 90089}

\date{\today}

\begin{abstract}

We study the random XY spin chain in a transverse field by
analyzing the susceptibility of the ground state fidelity, 
numerically evaluated through a standard mapping of the 
model onto quasi-free fermions. 
It is found that the fidelity susceptibility and its scaling properties provide useful information 
about the phase diagram. In particular it is possible to determine the Ising critical line 
and the Griffiths phase regions, in agreement with previous analytical and numerical results. 
\end{abstract}

\pacs{Valid PACS appear here}

\maketitle

{\em Introduction.--}
In the last few years concepts borrowed from quantum information theory 
have proven useful in characterizing the critical behavior 
of quantum many-body systems  \cite{AmFaOs}. 
In particular, a geometric approach to the study of quantum phase transitions (QPTs), i.e. 
the fidelity analysis, has been shown to be an effective way of characterizing 
distinct phases of quantum systems \cite{ZaPa, ZaGiCo, CVZ, YoLiGu, ZhBa, ZhZhLi, Zh}.  
Previously, the fidelity approach has been applied to a 
variety of homogeneous systems. In this work we extend 
these studies to disordered quantum systems. 
Specifically, 
we investigate the behavior of the fidelity susceptibility of the 
disordered XY model in a transverse field. 
It is well known that the presence of quenched disorder can have drastic effects on critical properties. 
The appearance of new universality classes and novel
states of matter such as the Griffiths phase are two important examples 
\cite{Fis1, Fis2, Gri, Sac}. 
The aim of the present work is to show what can be 
inferred about the physics of the disordered quantum system from the 
properties of the fidelity susceptibility.

The Hamiltonian of the disordered XY chain is given by
\begin{equation} \label{eq:Hamxy}
H=-\sum_{i=0}^{L-1} 
\left( \frac{1+\gamma_i}{2} \sigma_i^x \sigma_{i+1}^{x} +
\frac{1-\gamma_i}{2} \sigma_i^y \sigma_{i+1}^{y} + \lambda_i \sigma_i^z \right),
\end{equation}
where $\sigma_i^{\left\{x,y,z\right\}}$ are Pauli spin matrices, and 
$\gamma_i$ and $\lambda_i$ are 
sets of independent random coupling and field 
variables with distributions $\pi(\gamma_i)$ and 
$\rho(\lambda_i)$. 
Note that due to gauge symmetry the Hamiltonian (\ref{eq:Hamxy})
can be chosen to have only positive couplings and fields. 
This model can be mapped onto a system of quasi-free fermions with periodic boundary conditions, and 
an exact expression for the fidelity susceptibility is obtained
which depends explicitly on the random parameters characterizing the ground state of the system. 
In this work, we investigate the statistical properties 
of the fidelity susceptibility \cite{CVZ, YoLiGu} for relevant regions of parameter space.

Gaussian distributions are used for the random variables,
\begin{equation} \label{eq:gdistro}
\pi(x_i)=\rho(x_i) = \frac{1}{\sigma \sqrt{2 \pi}}
\exp\left\{{-\frac{1}{2}\left(\frac{x_i-x}{\sigma}\right)^2}\right\},
\end{equation}
where $x_i$ is either the field or the coupling at position $i$ on the chain, $x$ is the respective 
average value  and $\sigma^2$ is the variance.

{\em Previous Results.--}
The pure XY chain has been analytically solved in \cite{LiScMa}. In the absence of 
disorder two different quantum phase transitions are present. Following the standard 
notation, we refer to  the QPT driven by the transverse magnetic field $\lambda$ as the \textit{Ising transition}, 
 and to the QPT driven by the coupling parameter $\gamma$ as the \textit{anisotropy transition}. 
The Ising transition separates a ferromagnetic ordered phase  from 
 a paramagnetic quantum-disordered phase, whereas
the anisotropy critical line is the boundary between 
a ferromagnet ordered along the $x$ direction and a 
ferromagnet ordered along the $y$ direction.

A major improvement in the understanding of the 
effect of disorder on the physics of quantum magnets has 
been achieved with the use of the strong-disorder renormalization 
group technique (SDRG) by 
Dasgupta and Ma \cite{DasMa}, and further developed by Fisher \cite{Fis1, Fis2}. 
The correctness of this method has been 
corroborated both by numerics \cite{HaRiDa, YoRi} and 
analytic exact studies \cite{McK, BuMc}. 
In the work of McKenzie and Bunder \cite{McK, BuMc} the critical behavior of the 
disordered XY chain in a transverse field has been studied using 
a mapping to random-mass Dirac equations. 
The properties of the 
solutions of these equations imply the 
disappearance of the anisotropy transition in the presence of 
disorder. Furthermore, Griffiths phases are predicted to appear 
both around the Ising critical line and the anisotropy $\gamma=0$ line.

For $ \gamma=1 $ 
the XY random chain is closely related to the random transverse-field Ising chain
(RTFIC), which is a  prototypical model for disordered quantum systems. 
Since it is representative of the universality class of Ising transitions 
for all values of $ \gamma$, let us
briefly review what is known for this model. 
The Hamiltonian of the RTFIC is
$H=-\sum_{i=0}^{L-1} 
\left[J_i \sigma_i^x \sigma_{i+1}^{x} +
h_i \sigma_i^z 
\right],
$
where $J_i$ and $h_i$ are random 
couplings and fields respectively. The system is at criticality when 
the average value of the field equals the 
average value of the coupling. 
Using the SDRG one obtains that, at the quantum critical 
point, the time scale $\tau$ and the length scale $L$ are related by 
$\ln{\tau} \sim L^{1/2}$. This results in an infinite 
value for the dynamical exponent $ z $ at criticality. The distribution 
of the logarithm of the energy gap, $\ln{\epsilon}$, at criticality broadens
with increasing system size, in 
accordance with the scaling relation $\ln{\epsilon} \sim -L^{1/2}$ \cite{YoRi}. 
In the vicinity of the critical point 
the distribution of relaxation times is broad due to Griffiths singularities. 
This region of the parameter space, the Griffiths phase, 
is characterized by a dynamical exponent $z$ which 
depends on the distance from the critical point. This 
dependence is one of the hallmarks of the Griffiths phase. 

{\em Method.--}
The main idea of the fidelity approach is to detect QPTs 
through enhanced orthogonalization rates between 
ground states $| \Psi(x) \rangle $ nearby in parameter space. The orthogonalization 
is signaled by a drop in the fidelity, $ F(x,x + \Delta x) \equiv \left| \langle \Psi(x) | \Psi(x+\Delta x) \rangle \right |$, 
at the critical point. The fidelity susceptibility is a related quantity with a more transparent physical 
meaning  \cite{YoLiGu, CVZ}, and whose behavior is more suitable for numerical 
analysis. It is defined as
\begin{equation}\label{eq:chi}
\chi (x)=\lim_{\Delta x \rightarrow 0} \frac{-2 \ln{F(x,x + \Delta x)}}{\Delta x^2}.
\end{equation}
In  \cite{YoLiGu} it was shown that $\chi$ is related 
to the dynamic structure factor of the relevant operator 
associated with the transition. A generalization of this 
result, valid for the so-called \textit{geometric tensor}, 
has been given in \cite{CVZ}.

Previous works have characterized the pure XY spin chain 
using the fidelity approach \cite{ZaPa, ZaCoGi, ZaGiCo, CoGiZa} and the quantum Chernoff bound \cite{AbJaZa}. 
The mapping of the spin model onto the quasi-free fermion Hamiltonian \cite{LiScMa}, 
\begin{equation}\label{eq:Hamfree}
H=\sum_{i,j=1}^{L} c^{\dagger}_i A_{ij}c_j + 
\frac{1}{2} \sum_{i,j=1}^{L} \left( c^{\dagger}_i B_{ij}c^{\dagger}_j + H.c.\right),
\end{equation}
yields an explicit BCS-like form for the 
ground state
$\vert \Psi \rangle= \mathcal{N} \exp{\left( \frac{1}{2} \sum_{j,k=1}^{L} c_j^{\dagger} G_{jk} c_k^{\dagger} \right)} 
\vert 0 \rangle,$
where $ \mathcal{N} $ is a normalization factor.

The fidelity of the ground states evaluated 
at slightly different parameter values (coupling or magnetic field)
$ x $ and $ x + \Delta x $ has a simple analytical expression. 
Defining the matrix $Z(x) \equiv A(x)-B(x)$ and the unitary part
of the polar decompositions of $Z(x)$ and $\tilde{Z} \equiv Z(x + \Delta x)$ 
as $T(x)$ and $\tilde{T} \equiv T(x + \Delta x)$, respectively, the fidelity can be written as
\begin{equation}\label{eq:fidedet}
F(Z,\tilde{Z})=\sqrt{\vert \det{\frac{T+\tilde{T}}{2}} \vert}.
\end{equation}
Note that the matrix $G$ defining the ground state is simply 
the Cayley transform of $T$ \cite{CoGiZa}.

In the following, we will use an alternative
expression for the fidelity susceptibility, obtained in the limit of small $ \Delta x $,
\begin{equation}\label{eq:chifrob}
\chi (x) =\frac{1}{8}\Vert\partial_{x}T \Vert_{F}^{2},
\end{equation}
with  $\Vert \cdot \Vert_{F}$ the Frobenius norm. 
Eq. (\ref{eq:chifrob}) is obtained from (\ref{eq:fidedet}) via standard algebra. 
We have numerically evaluated the fidelity susceptibility 
using (\ref{eq:chifrob}) for relevant regions of parameter space of 
the disordered XY model. 
The numerical analysis has been performed on two sets of system sizes, i.e. 
\{128, 256, 512\}, and \{400, 410, ...,  500\} in steps of 10. 
We have taken 
50,000 disorder realizations for all sizes except for those larger 
than 400, in which case we used 10,000 realizations.

{\em Results.--}
We consider the Hamiltonian (\ref{eq:Hamxy}), where 
the couplings $\gamma_i$ and 
the transverse fields $\lambda_i$ are independent random variables 
with Gaussian distributions centered around $\lambda \equiv \left[ \lambda_i \right]_{\textrm{ave}}$ 
and $\gamma \equiv \left[ \gamma_i \right]_{\textrm{ave}}$, both with 
standard deviation $\sigma$. 
$\left[ \cdot \right]_{\textrm{ave}}$ denotes the arithmetic mean over the disorder realizations.

A scaling analysis has been performed using arguments first developed 
in \cite{CVZ}. 
Following that reference, we can express the fidelity susceptibility as 
an integral in imaginary time 
$\chi=\int_{-\infty}^{\infty} d \tau \tau G(\tau),
$
where
$G(\tau)=\theta(\tau)\ll \partial_{x} H(\tau) \partial_{x}H(0) \gg
$
is the connected correlation function of the conjugate operator 
in the Hamiltonian  associated with the driving parameter in the 
transition, and $ \theta $ is the Heaviside step function. For example, in the case of the 
Ising transition we have
$G(\tau)=\theta(\tau)\ll \sum_{i,j} \sigma_i^z(\tau) \sigma_j^z(0) \gg$.
The average fidelity susceptibility can then be written as
$\left[ \chi \right]_{\textrm{ave}}= \int_{-\infty}^{\infty} d \tau \tau \left[ G(\tau) \right]_{\textrm{ave}} \sim L^{\Delta_{\chi}}$, 
where the finite-size scaling dimension $\Delta_{\chi}$
of $\left[ \chi \right]_{\textrm{ave}}$ is given by
$\Delta_{\chi}=2 z+2 - 2 \Delta_O$ \cite{CVZ}.
$ \Delta_O $ is the scaling dimension of the QPT conjugate operator ($ \sum_i \sigma_i^{z} $ in the 
case of the transition driven by $ \lambda $), and in 
general $\Delta_O$ depends on the parameters $\gamma$ and $\lambda$. 

For the XY chain without disorder, in the quantum critical regions $\chi$ scales
as $\chi \sim L^2$, whereas
away from the critical region $\chi \sim L$.  Since for finite system sizes the 
quantum critical region has a finite width, $\Delta_{\chi}$ is $1$ for all but 
a narrow range of $\lambda$ (or $\gamma$), having a maximum of $2$ for $\lambda = 1$ ($\gamma=0$).

With this disorder-free behavior in mind, 
we now study $\left[ \chi \right]_{\textrm{ave}}$ about the Ising transition,
driven by the coupling $\lambda$.
In our numerical studies we have focused on the case of the RTFIC, where $\gamma = 1$.  Qualitatively
all of our results on the critical behavior of the fidelity susceptibility hold true for other values of $\gamma$, since 
the universality class of the model does not change in the range
$\gamma \in \left( 0,1 \right]$.

\begin{figure}[htp]
\includegraphics[scale=0.3]{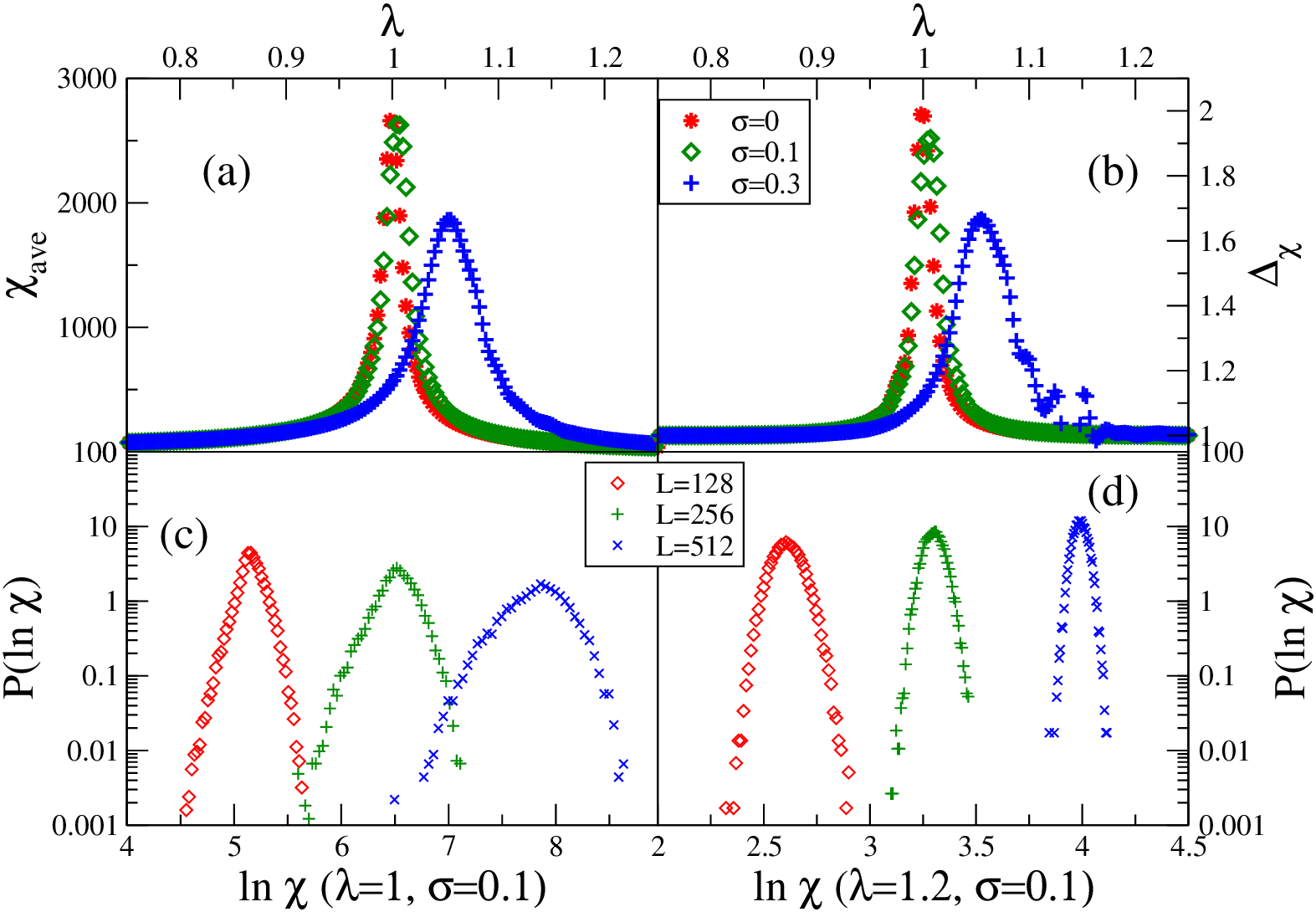} 
\caption{Ising transition at $\gamma = 1$.
(a) Average fidelity susceptibility, $\left[ \chi \right]_{\textrm{ave}}$, for $L=512$ and $10^4$ realizations, with varying 
disorder strengths $\sigma \in \{0,0.1,0.3\}$,
(b) the associated finite-size scaling dimension $\Delta_{\chi}$ of $\left[ \chi \right]_{\textrm{ave}}$, 
(c) probability distribution of $\ln \chi$ at the Ising transition for system sizes $L=128,256,512$ and disorder $\sigma = 0.1$, 
(d) distribution of $\ln \chi$ away from the Ising transition for the same disorder and range of system sizes.}
\label{fig:Ising4x4}
\end{figure}

Fig. \ref{fig:Ising4x4}(a) shows $\left[ \chi \right]_{\textrm{ave}}$ as a function of $\lambda$ for the clean 
case and $\sigma = 0.1, 0.3$ disorder strengths.
The averaged fidelity susceptibility displays a local maximum at the Ising critical point, which for the disordered 
case is shifted slightly away from the clean value of $\lambda =1$ due to finite size effects.
Fig. \ref{fig:Ising4x4}(b) shows $\Delta_{\chi}(\lambda)$, the finite-size scaling dimension of $\left[ \chi \right]_{\textrm{ave}}$, 
for the same set of disorder strengths. The disorder leads to a broadening in the peak of $\Delta_{\chi}$, which is consistent 
with the presence of a Griffiths phase.
Note that far from the Ising critical point $\left[ \chi \right]_{\textrm{ave}}$ scales strictly extensively, while in the vicinity 
of the critical point the scaling becomes superextensive. 
For the weaker noise this scaling is nearly quadratic, as in the clean case, while with stronger noise the 
maximum scaling dimension is correspondingly reduced. 
Qualitatively, the reduction of the maximum scaling dimension may be ascribed to the presence of rare regions whose 
extent effectively determines the critical behavior.  The linear extension of rare regions is smaller than the 
overall system size determining the critical behavior in the clean case.\\ \\

In Fig. \ref{fig:Ising4x4}(c) and (d) we plot the distribution of the 
fidelity susceptibility over many 
realizations at the Ising critical point and away from it, for system sizes $L=128, 256$, and $512$. We choose to 
plot the distribution of $\ln \chi$ instead of $\chi$ itself because, in analogy to other physical quantities, 
the presence of disorder greatly broadens the distribution. 
As the system size increases, note that the probability density function of $\ln \chi$ broadens for $\lambda=1$, but 
becomes narrower away from criticality. Indeed, this broadening behavior persists for a range of values of 
$\lambda$ about the critical point.
This is typical of disordered systems, and is analogous to the absence of self-averaging of some physical observables.

The Griffiths phase around the Ising critical point can be detected by looking at the scaling dimension of the fidelity 
susceptibility and at the properties of the distribution of $\ln \chi$, in accordance with the relation 
$\Delta_{\chi}=2 z+2 - 2 \Delta_O$.  The following analysis of the region about 
the anisotropy line further supports this conclusion.

\begin{figure}[htp]
\includegraphics[scale=0.3]{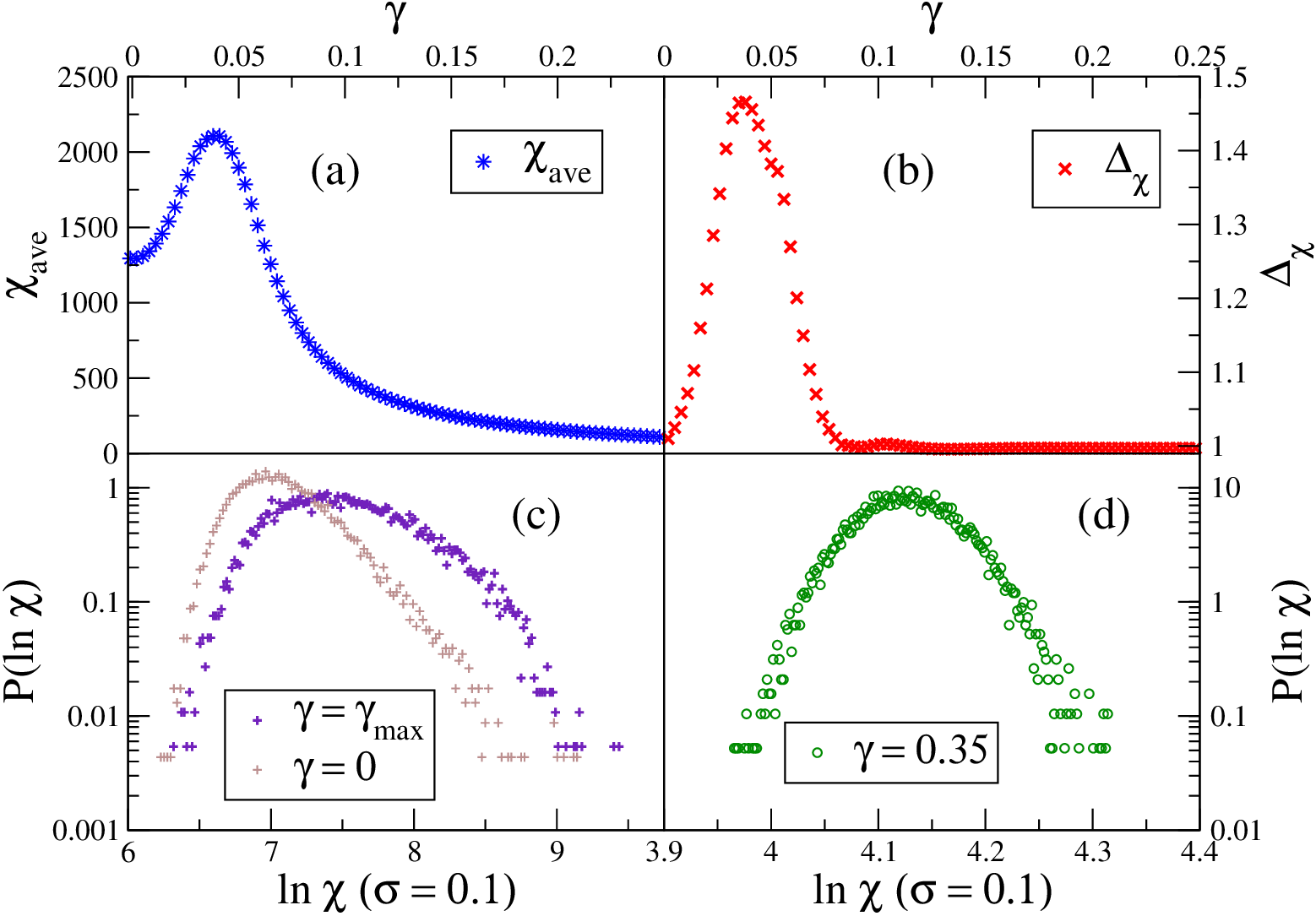} 
\caption{Elimination of anisotropy transition at $\gamma=0$ due to 
disorder. (a) Average fidelity susceptibility $\left[ \chi \right]_{\textrm{ave}}$, near $\gamma = 0$ for $L=500$, $\lambda=0.2$, $\sigma = 0.1$, and $10^4$ realizations. 
(b) Finite-size scaling dimension of $\left[ \chi \right]_{\textrm{ave}}$ (in this case $\gamma = 0.036$), 
(c) probability distribution 
of $\ln \chi$ at $\gamma=0$ and at the value of $\gamma$ corresponding to the maximum of $\left[ \chi \right]_{\textrm{ave}}$, (d) distribution 
of $\ln \chi$ far away from the anisotropy line, where the finite-size scaling is extensive.}
\label{fig:Ani4x4}\underline{}
\end{figure}

\begin{figure}[htp]
\centering
\includegraphics[scale=0.3]{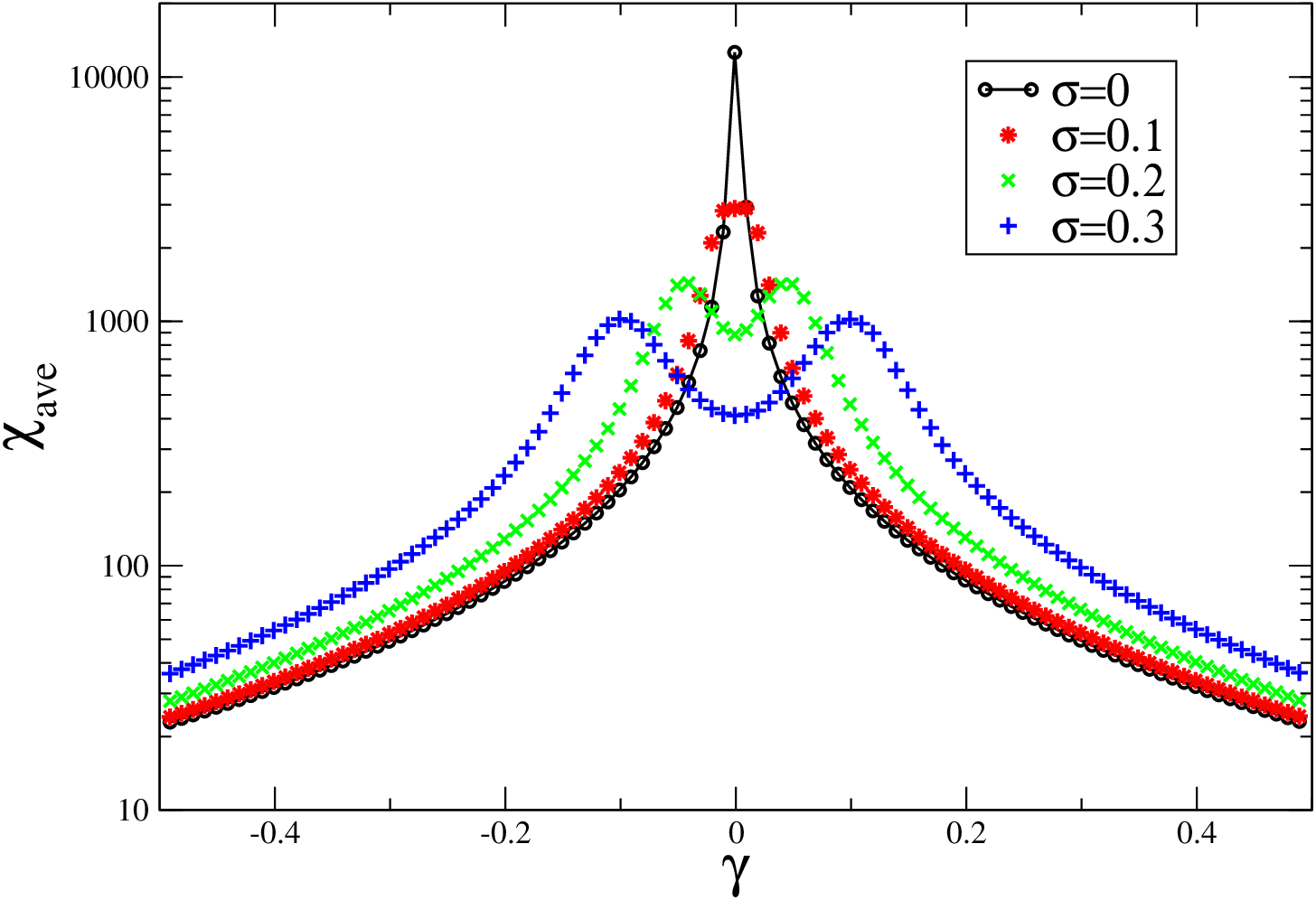} 
\caption{Average fidelity susceptibility for various disorder strengths $\sigma \in \{0,0.1,0.2,0.3\}$, with fixed system size $L=400$. 
Here $\lambda=0.5$, and the derivative in Eq. (\ref{eq:chifrob}) is taken along $\gamma$.}
\label{fig:AniVariousDisorder}
\end{figure}

Although for the disordered XY model the $\gamma=0$ line is 
not critical, as it is in the pure case \cite{McK, BuMc}, 
the presence of Griffiths singularities 
still has highly non-trivial effects on the fidelity susceptibility 
in the vicinity of $\gamma=0$, as
shown in Fig. \ref{fig:Ani4x4}(a). Specifically, in the presence of 
disorder the peak in $\left[ \chi \right]_{\textrm{ave}} (\gamma)$ splits into two peaks, 
symmetrical about $\gamma=0$. Note that $\gamma = 0$ is a special
case for the XY chain.  With zero anisotropy and noise only in the 
field $\lambda_{i}$, Bunder and McKenzie \cite{BuMc} showed that the density of states 
does not diverge at zero energy.  This implies that this point is not critical and does not belong to 
a Griffiths phase.  
We observe similar behavior with disorder in both the field and anisotropy. 
At $\gamma = 0$ the fidelity susceptibility scales only extensively, as it does in the 
non-critical region.
This suggests that the $\gamma = 0$ point does not belong to the Griffiths phase, having 
characteristics of a point which is non-critical and away from any Griffiths phase. 
This is further corroborated by the non-monotonic dependence on $\gamma$ of the associated 
scaling dimension shown in Fig. \ref{fig:Ani4x4}(b). At 
$\gamma = 0$, one finds $\Delta_{\chi} = 1$, whereas in the interval
$0 < \gamma < 0.075$ the scaling dimension $\Delta_{\chi} (\gamma )$ exhibits a 
non-universal dependence on the driving parameter $\gamma$,
indicating the presence of a Griffiths regime.
Note that the observed maximum is not to be seen 
as an indication of a QPT.  Rather, it originates from 
the competition between the scaling properties of $ \chi $ in the 
Griffiths phase and at the $ \gamma=0 $ line. 
In Figs. \ref{fig:Ani4x4}(c) and (d), 
we show $P(\ln \chi )$ at $\gamma = 0$, at the point where $\chi (\gamma)$ and $\Delta_{\chi}(\gamma )$ both peak,
and far away from the anisotropy line. In analogy to 
the Ising transition, the probability distribution function in the 
Griffiths regime is broad and asymmetric due to absence of self-averaging, whereas far away from it its
shape is symmetric and its distribution is much narrower.
To complete the discussion of the effects of disorder on the anisotropy 
transition in Fig. \ref{fig:AniVariousDisorder} we plot the average fidelity susceptibility for 
a fixed system size and various disorder strengths, including the clean case. 
Notice 
that as the disorder strength is increased, the original peak 
disappears and the new maxima in the fidelity susceptibility 
are symmetrically located around the $\gamma=0$ line, at a 
distance which increases with disorder. Much like 
the Ising transition, the maximum value of the fidelity susceptibility 
decreases with increased disorder. We believe that 
this can be explained again in terms of the 
extension of rare regions.

{\em Conclusions.--}
In this work we have applied the fidelity approach to the 
study of the disordered XY chain in an external magnetic field. 
We have found that the fidelity susceptibility is able to 
provide the phase diagram for this model. In the case of the Ising transition, 
we obtain results which are consistent with what is already known in the literature. 
In the parameter region around the $\gamma=0$ line the scaling 
analysis of the fidelity susceptibility shows the disappearance 
of the QPT and the emergence of a Griffiths phase, in accordance 
with similar analytical and numerical results. As far as we know, this 
result has not been obtained before for this distribution of disorder 
both in the couplings and in the fields. This is nontrivial, 
since it is known that choosing a different parametrization for the disorder 
can modify the critical behavior \cite{McK}.

We plan 
to further investigate the relevance of disorder on the fidelity susceptibility in future works. 
Other aspects that will be studied with more details are the extent 
of the Griffiths phase together with its dependence on disorder strength and the 
probability distribution of disorder.

We thank H. Saleur and L. Campos Venuti for helpful discussions. 
Computation for the work described in this paper was supported by the 
University of Southern California Center for High Performance Computing 
and Communications. We acknowledge financial support by the National 
Science Foundation under grant DMR-0804914.

\end{document}